\documentclass[pre,letterpaper,twocolumn,showpacs]{revtex4}

\usepackage{epsfig}
\usepackage{graphicx}
\usepackage{amssymb}

\begin{document}

\title{Lifetime of dynamic heterogeneities in a binary Lennard-Jones mixture}

\author{Elijah Flenner}
\author{Grzegorz Szamel}
\affiliation{Department of Chemistry, Colorado State University, Fort Collins, CO 80525}

\begin{abstract}
A four-time correlation function was calculated using a computer 
simulation of a binary Lennard-Jones mixture.
The information content of the four-time correlation function is similar to that of
four-time correlation functions measured in NMR experiments. The correlation function selects a 
sub-ensemble and analyzes its dynamics after some waiting time. 
The lifetime of the sub-ensemble selected by the four-time correlation
function is calculated, and compared to the lifetimes of slow sub-ensembles
selected using two different definitions of mobility, and to the $\alpha$ relaxation time.
\end{abstract}

\date{\today}


\maketitle

The origin of the non-exponential relaxation found in 
supercooled liquids has been studied extensively in the last ten years.
Two possibilities exist \cite{Ediger,Richert}.  Either all the particles 
undergo non-exponential relaxation (homogeneous scenario),
or the relaxation of each particle is exponential and there is a large variation in the
relaxation time of the particles (heterogeneous scenario). 
There have been many simulations \cite{KobDH,Qianrate,Qian,Heuersim,Donati,Doliwa,GlotzerDH,Lacevic,Lacevic2} and 
experiments \cite{Bohmer,Cicerone,Wang,Blackburn,Schmidt} which imply heterogeneous relaxation.  
The heterogeneous relaxation scenario suggests that
the particles in a supercooled liquid can be categorized by 
their relaxation time.  
The particles with the shortest relaxation times are referred to as ``fast'' particles, and the 
particles with the longest relaxation times are ``slow'' particles.
One important question is the lifetime of the dynamic heterogeneities, \textit{i.e.} how long does a fast
particle remain fast and a slow particle remain slow? The first part of this question was 
considered 
in one of the early simulational investigations of dynamics heterogeneities \cite{KobDH}: the lifetime
of fast particles has been found to be much shorter than the $\alpha$ relaxation time.
It should be noted that experiments are usually sensitive to slow particles and thus 
simulational investigation of the slow particles lifetime is also important; however,
to the best of our knowledge, lifetime of slow particles 
has been studied only
in two dimensions where it has been found to be comparable to the $\alpha$ relaxation time 
\cite{Perera}.
Here we study the lifetime of slow particles using an approach inspired by
one of the experimental protocols.
Our study is complementary to recent investigations of the
spatial correlations of the slow particles \cite{Lacevic,Lacevic2}.

The lifetime of dynamic heterogeneities has been measured in 
a reduced four-dimensional nuclear magnetic resonance (NMR) experiment
by monitoring parts of a four-time correlation function. The general idea of the 
experiment has been lucidly explained by Heuer \cite{Heuer}: one can define a filtering function 
$f(t_1,t_2)$ such that $\left<f(t_1,t_2)\right>$ selects 
particles which are slow over a time interval $\Delta t_{12} = t_2 - t_1$.   
Thus, $\left<f(t_1,t_2)f(t_3,t_4)\right>$
selects particles which are slow over time intervals $\Delta t_{12}$ 
and $\Delta t_{34} = t_4 - t_3$.  The two time intervals 
are separated by a waiting time $t_w = t_3 - t_2$.
For small $t_w$, the relaxation of the slow 
sub-ensemble remains slow, but for large enough $t_w$ the relaxation of the 
slow sub-ensemble is the same as the relaxation of the 
full ensemble. The lifetime of the slow ensemble is related to the minimum $t_w$ such that 
the average relaxation time of the slow sub-ensemble returns to the average relaxation time of the 
full ensemble. B\"ohmer 
\textit{et al.}~\cite{Bohmer} used this idea to investigate Ortho-Terphenyl (OTP) 
at 10 K above $T_g = 243$~K.  
Using a pulse sequence they selected a set of particles which did not rotate appreciably 
over a time interval $\Delta t_{12}$, \textit{i.e.}\ a slow sub-ensemble. 
The particles were then allowed
to evolve during a time interval $t_w$. Finally they measured what 
fraction of the slow sub-ensemble were still slow over a time interval $\Delta t_{34}$.   
The characteristic time for the slow sub-ensemble to remain slow 
was found to be comparable to the average relaxation time of the full ensemble.  
This is in a stark contrast with results obtained for OTP by Ediger's group\cite{Cicerone,Wang}: 
at $T_g+4$ K  the lifetime of the dynamic heterogeneities was found to be 6 times
longer than the $\alpha$ relaxation time and at 
$T_g + 1$ K it was 100 times longer! Ediger's findings could, however, be compatible with
the NMR result if strong temperature dependence of the lifetime sets in close to $T_g$.  

The procedure used in this work to measure the lifetime of dynamic heterogeneities is closely
related to the NMR approach described above.  We use a four-time correlation function 
to select a slow sub-ensemble, and monitor the relaxation and the 
lifetime of the slow sub-ensemble. 
The four-time correlation function selects a sub-ensemble without any explicit
definition of mobility, thus it is not clear which particles are 
contributing to the four-time correlation function.
To identify these particles we use different definitions of mobility 
to select sub-ensembles whose relaxation is similar
to the sub-ensemble selected by the four-time correlation function. Finally, we
measure the lifetime of these slow sub-ensembles.

\begin{figure}[t,b]
     	\includegraphics[scale=0.29]{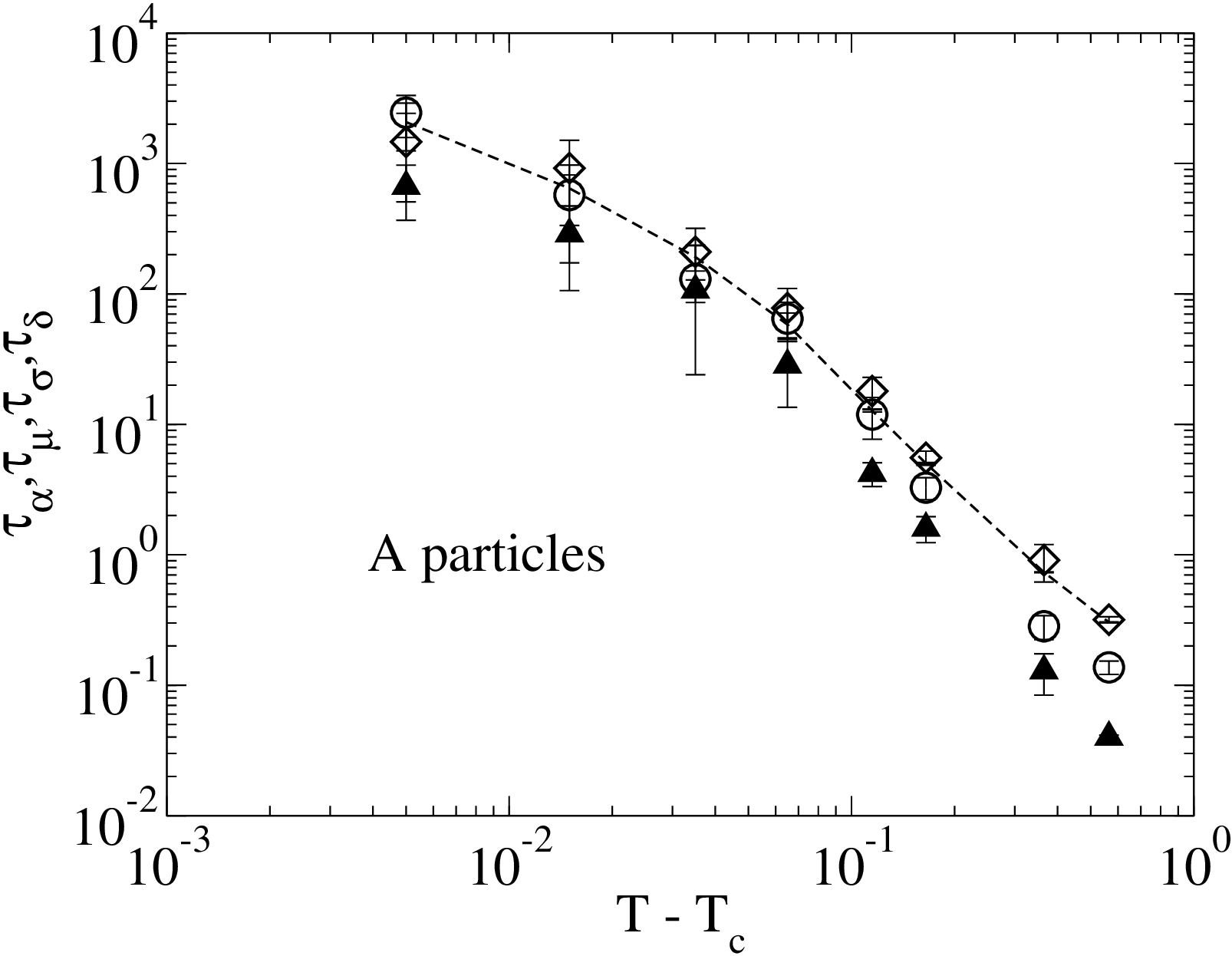}\\[24pt]
	\includegraphics[scale=0.29]{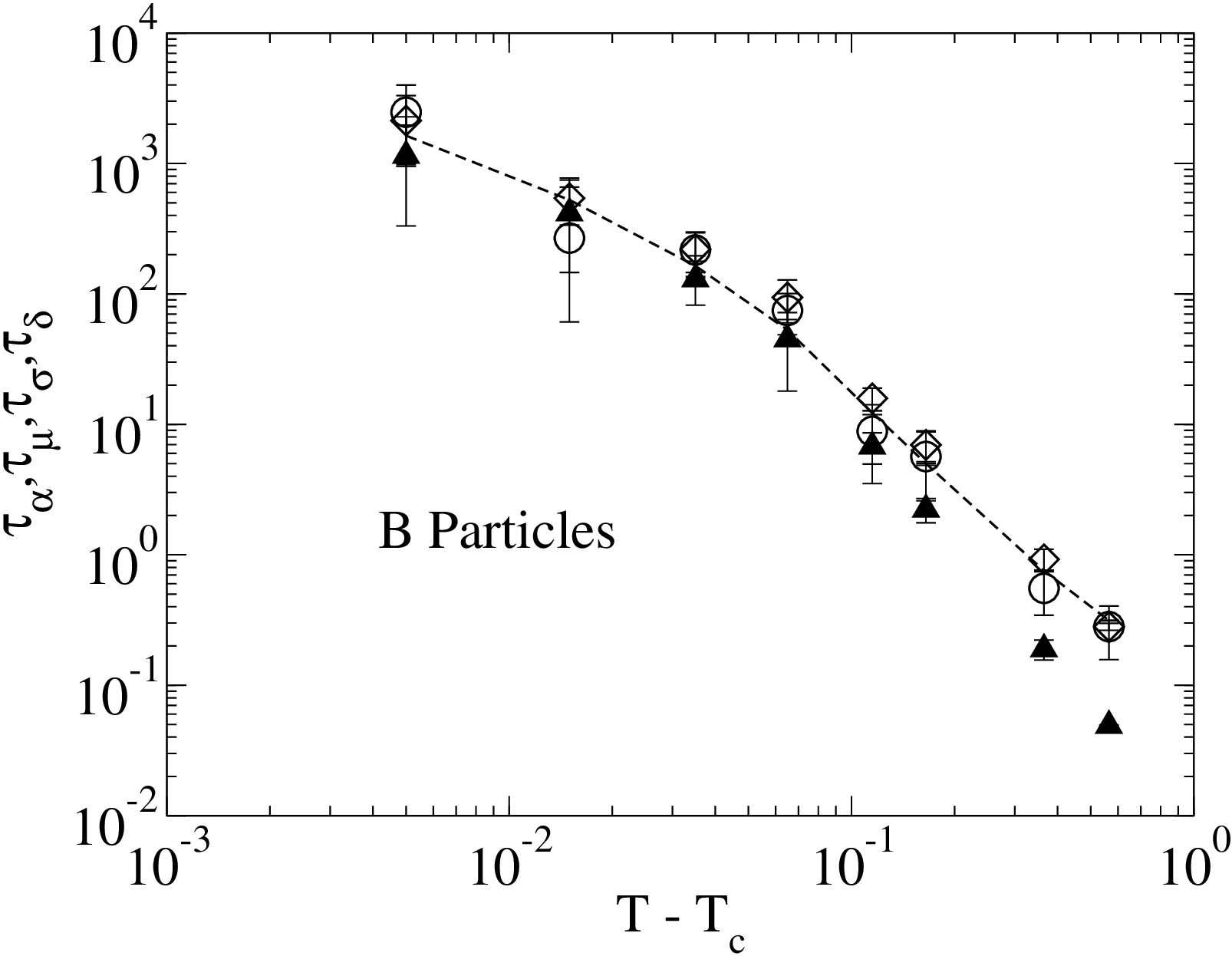}
\caption{The characteristic lifetime found using the four-time
correlation function $(\blacktriangle)$, by using $\sigma_i$ (Eq. (\ref{s}))
to define the mobility $(\Diamond)$, and by using $\delta_i$ (Eq. (\ref{d}))
to define the mobility $(\circ)$, compared to the $\alpha$-relaxation
time (dashed line).\label{lifetime}}
\end{figure}

To investigate the lifetime of dynamic heterogeneities we use the trajectories generated
by an extensive Brownian Dynamics simulation study of a 80:20 mixture of 
a binary Lennard-Jones fluid \cite{KobLJ}.  
Briefly, the potential is given by 
$V_{\alpha \beta} =  4\epsilon_{\alpha \beta} 
\left[ (\sigma_{\alpha \beta}/r)^{12} - (\sigma_{\alpha \beta}/r)^6 \right]$,
where $\alpha,\beta \in \{A,B\}$, and $\epsilon_{AA} = 1.0$, $\epsilon_{AB} = 1.5$,
$\epsilon_{BB} = 0.5$, $\sigma_{AA} = 1.0$, $\sigma_{AB} = 0.8$, and $\sigma_{BB} = 0.88$.  
A total of $N = N_A + N_B = 1000$ particles were simulated with a fixed cubic box length of $9.4\sigma_{AA}$.  
All the results are presented in reduced units where $\sigma_{AA}$ and $\epsilon_{AA}$
are the units of length and energy, respectively.  The system was simulated at temperatures
$T = 0.44$, 0.45, 0.47, 0.5, 0.55, 0.6, 0.8 and 1.0.  A long equilibration run, and
two to eight production runs were performed 
at each temperature. The equilibration run was at least as long as the production runs.
The presented results are the average of the production runs.
The characteristics of this glass-forming liquid has been extensively studied \cite{KobLJ,Gleich,KobCM}.  
The details and the results of the 
Brownian dynamics simulation are given elsewhere \cite{SzamelFlenner}. 
In particular, we found that $\alpha$ relaxation times, Fig.~\ref{lifetime}, follow a power-law
temperature dependence in the temperature range $0.47\le T \le 0.8$ and deviate from this power-law
dependence for $T<0.47$. This is similar to earlier findings using Newtonian \cite{KobLJ,KobCM} 
and stochastic dynamics \cite{Gleich}.

To examine lifetime of dynamic heterogeneities we follow the procedure 
discussed above: we use a filtering function 
%$f(t_1,t_2) = \exp \left[i\mathbf{q}\cdot\left(\mathbf{r}_j(t_2) - \mathbf{r}_j(t_1) \right) \right]$,
$f(t_1,t_2) = e^{i\mathbf{q}\cdot\left(\mathbf{r}_j(t_2) - \mathbf{r}_j(t_1) \right)}$,
where $\mathbf{r}_j(t)$ is the position of particle $j$ at time $t$. Thus $\left<f(t_1,t_2)\right>$
is the incoherent intermediate scattering function $F_s(q;t_2-t_1)$. 
For all the calculations, $q$ is set to a value around the first peak in the $AA$ ($q = 7.25$) 
or $BB$ ($q = 5.75$) partial structure factor
for $\mathcal{M}^A$ and $\mathcal{M}^B$, respectively.
The four-time correlation function is defined as follows:
\begin{eqnarray}
\nonumber
\lefteqn{\mathcal{M}^{\alpha}(q,t_1,t_2,t_3,t_4)  =
 \frac{\left<f(t_1,t_2)f(t_3,t_4)\right>}{\left<f(t_1,t_2)\right>}} \\
 = &  \frac{\displaystyle
\left<\frac{1}{N_{\alpha}}
\sum_{j=1}^{N_{\alpha}}
e^{i\mathbf{q}\cdot\left(\mathbf{r}_j(t_2) - \mathbf{r}_j(t_1)\right)}
e^{i\mathbf{q}\cdot\left(\mathbf{r}_j(t_4) - \mathbf{r}_j(t_3)\right)}
\right>}
{\displaystyle
\left<\frac{1}{N_{\alpha}} \sum_{_j=1}^{N_{\alpha}}
e^{i\mathbf{q}\cdot\left(\mathbf{r}_j(t_2) - \mathbf{r}_j(t_1)\right)}
\right>}
\end{eqnarray}
where $\alpha \in \{A,B\}$.
The normalization of the correlation function is such that if $t_3 = t_4$, then $\mathcal{M}^{\alpha}$ = 1.0.
For small $t_w=t_3 - t_2$, the relaxation of the slow 
sub-ensemble remains slow, but for large enough $t_w$ the relaxation of the 
slow sub-ensemble is the same as the relaxation of the 
full ensemble. 

\begin{figure}
      	\includegraphics[scale=0.29]{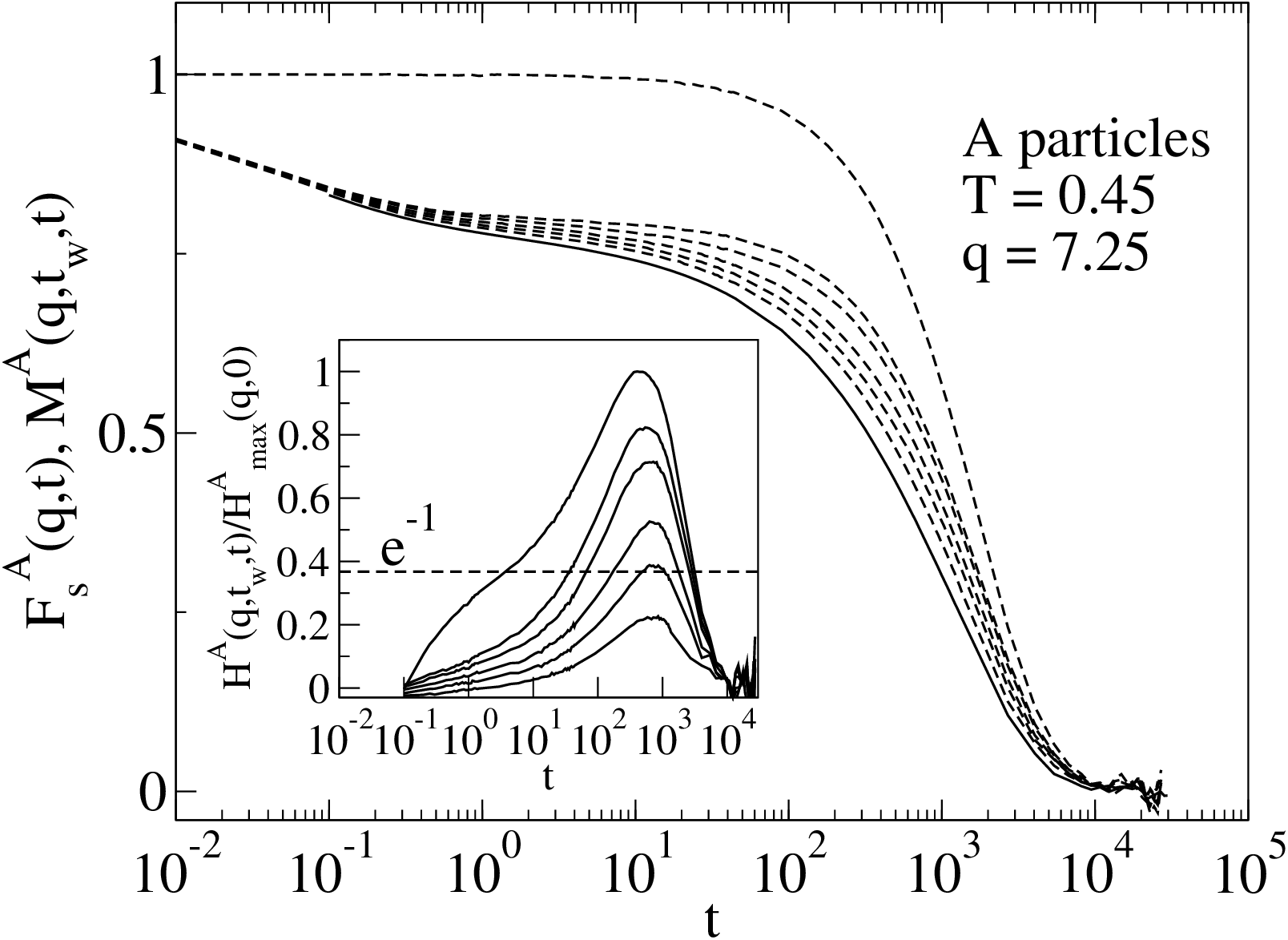}
\caption{$F_s^A(q,t)$ (solid line) and $M^A(q,t_w,t)$ (dashed lines) for 
$t_w = 0$, 5, 50, 250, 500, and 1000 at $T = 0.45$ listed in order from the longest relaxation
time to the shortest relaxation time.  (Insert) $H^A(q,t_w,t)/H^A_{max}(q,0)$ for 
$t_w = 0$, 5, 50, 250, 500, and 1000 at $T = 0.45$.\label{fourtime}}
\end{figure}

We fix the first time interval,  $\Delta t_{12} = t_2 - t_1$, to be equal to 
$3\tau_{\alpha}$ where $\tau_{\alpha}$ is the $\alpha$ relaxation time
($\tau_{\alpha}$ is defined by the usual relation $F_s(q,\tau_{\alpha}) = e^{-1}$).
This is comparable to the longest time intervals $\Delta t_{12}$
used to select a slow sub-ensemble in the NMR experiment of B\"ohmer \textit{et al}. Note that  
the time $\Delta t_{12} = 3\tau_{\alpha}$ 
is well past the plateau region of the mean squared displacement, and is longer
than what has been used in previous simulational investigations 
which examined dynamic heterogeneities \cite{KobDH,Qianrate,GlotzerDH}.
The second time interval, the waiting time $t_w = t_3 - t_2$, is varied. Finally, for a given $t_w$,
$M^{\alpha}(q,t_w,t) \equiv \mathcal{M}^{\alpha}(q,0,3\tau_{\alpha},3\tau_{\alpha}+t_w,t+3\tau_{\alpha}+t_w)$ 
is calculated as a function of time $t$ (\textit{i.e.} as a function of the last
time interval, $\Delta t_{34} = t_4 - t_3$).  $M^{A}(q,t_w,t)$
is shown in Fig.~\ref{fourtime} for several waiting times.  Notice that if 
$t_w = 0$, then $M^{\alpha}(q,t_w,t)=F_s^{\alpha}(q,3\tau_{\alpha}+t)/F_s^{\alpha}(q,3\tau_{\alpha})$.  
Also, $M^{\alpha}(q,t_w,t)$ converges to $F_s^{\alpha}(q,t)$ as the
waiting time increases.  The lifetime of the sub-ensemble measures
how long it takes for this convergence to occur.  

We define the lifetime of dynamic heterogeneities as the waiting time for which the difference
between $M^{\alpha}(q,t_w,t)$ and $F_s^{\alpha}(q,t)$ is equal to $e^{-1}$ of its value at 
short times. The exact procedure is as follows:
As shown in Fig.~\ref{fourtime}, for $t_w > 0$ there is an initial decay of $M^{\alpha}$ to a plateau region, then
$M^{\alpha}$ decays to zero after the plateau.  This is in contrast to the $t_w = 0$ case where
there is no initial decay to a plateau.
Since we are interested in the relaxation after the plateau, 
$M^{\alpha}(q,0,t)$ is multiplied by a temperature dependent
factor $C(T)$ so that $C(T)M^{\alpha}(q,0,t_c) = F_s^{\alpha}(q,t_c)$ where $t_c$ 
is at the beginning of the plateau region of $F_s^{\alpha}$.  The choice of $t_c$ affects 
the results slightly, with a larger $t_c$ leading to a somewhat longer lifetime.  However, the choice of
$t_c$ does not affect any of the conclusions of this work. We calculate 
\begin{equation}
H^{\alpha}(q,t_w,t) = \left\{ \begin{array}{ll}
C(T)M^{\alpha}(q,t_w,t) - F_s^{\alpha}(q,t) & t_w = 0 \\
M^{\alpha}(q,t_w,t) - F_s^{\alpha}(q,t) & t_w > 0
\end{array} \right. 
\end{equation} 
and determine the lifetime as the waiting time when the peak value of $H^{\alpha}(q,t_w,t)$, 
is a factor of $e$ smaller than its $t_w=0$ value,
\textit{i.e.}\ $H_{max}^{\alpha}(q,\tau_{\mu})/H_{max}^{\alpha}(q,0) = e^{-1}$,
where $H^{\alpha}_{max}(q,t_w)$ is the maximum value of $H^{\alpha}(q,t_w,t)$.  

Shown in Fig.~\ref{lifetime} is the temperature dependence of the lifetime $\tau_\mu$ of the 
slow sub-ensemble selected by $\mathcal{M}^\alpha$ and for comparison
the $\alpha$ relaxation time.  Notice that the lifetime is not longer than the $\alpha$ relaxation time.
The lifetime increases faster with decreasing temperature than the $\alpha$ relaxation
time except at the lowest temperatures studied where it 
has the same temperature dependence as the $\alpha$ relaxation time. 

An advantage of a computer simulation is that the trajectories of 
individual particles can be followed throughout the simulation.
This allows us to try to identify 
a slow sub-ensemble which is a major contribution 
to the four-time correlation function, \textit{i.e.}\ the sub-ensemble selected by $f(t_1,t_2)$.
To this end we have defined the mobility $\sigma_i(\Delta t)$ of a 
particle $i$ over a time interval $\Delta t$ as:
\begin{equation}\label{s}
\sigma_i(\Delta t) \equiv \overline{\left| \mathbf{r}_i(t) - \mathbf{r}_i(t_1) \right|^2},
\end{equation}
where the bar denotes an average over time $t \in (t_1,t_1+\Delta t)$ \cite{vollmayrlee}.
A particle is defined as slow over a time 
interval $\Delta t$ if $\sigma_i(\Delta t)$ is less than a cutoff value $r_{cut}^2$.
These are the particles which stay closest to their position 
at $t_1$ during the whole time interval $\Delta t$.  

\begin{figure}
      	\includegraphics[scale=0.29]{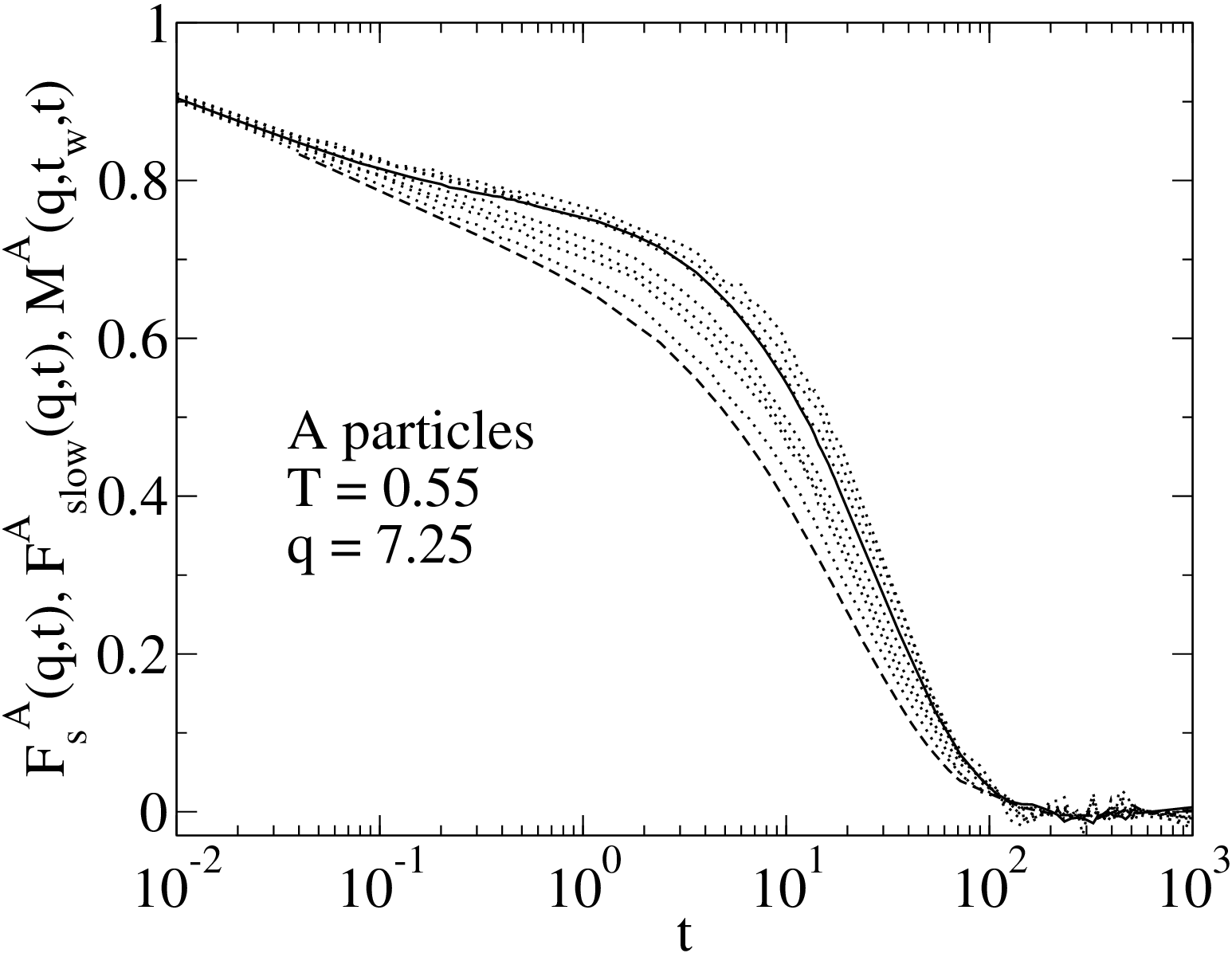}
\caption{$F_s^A(q,t)$ (dashed line), $F_{slow}^A(q,t_w,t)$ (dotted lines) for $r_{cut}^2 = 0.05$, 0.03,
0.025, 0.02, 0.015, 0.014, 0.013 listed from left to right, 
and $M^A(q,t_w,t)$ (solid line) for $t_w = 0.2$.\label{slow}}
\end{figure}

To make a connection with the four-time correlation function study we fix $\Delta t=3\tau_{\alpha}$.
Next, the incoherent intermediate scattering function, $F_{slow}^{\alpha}(q,t)$,
is calculated for the slow particles
after a waiting time $t_w$ has elapsed. $F_{slow}^A$ is shown in Fig.~\ref{slow}
for different values of $r_{cut}^2$, and is compared to $F_s^A$ and $M^A$. Note that $F_{slow}^A$
and $M^A$ are calculated for the same waiting time $t_w = 0.2$.  
%Note that the time interval used to identify the mobility of the particles is not
%used in the calculation of $F_{slow}^{\alpha}$. 
For a large cutoff $r_{cut}^2$, the sub-ensemble behaves like the full ensemble.  
For smaller values of $r_{cut}^2$, the average relaxation time of the slow particles 
is longer than the average relaxation time of the full ensemble.
For a small enough cutoff, $F_{slow}^{\alpha}(q,t) \approx M^{\alpha}(q,t_w,t)$.
The size of the cutoff needed to achieve this equality depends on $t_w$ and the time
interval used to identify the slow particles.
For the temperature shown in Fig.~\ref{slow} $F_{slow}^A(q,t) \approx M^A(q,0.2,t)$ for $r_{cut}^2 = 0.015$.  
This cutoff corresponds to the 0.075\% slowest particles.
As $t_w$ increases, the value of $r_{cut}^2$ resulting in $F_{slow}^{\alpha} \approx M^{\alpha}$ also increases. 
For the higher temperatures, it was not possible to find a value of $r_{cut}^2$ so that 
$F_{slow}^{\alpha} \approx M^{\alpha}$ for short waiting times.

The characteristic lifetime of the slow particles $\tau_{\sigma}$ can be calculated using the algorithm
described above (note that now we do not need the correction factor $C(T)$).  
The temperature dependence of the characteristic lifetime of the slow sub-ensemble is shown 
in Fig.~\ref{lifetime}. % for the $A$ and $B$ particles. 
The cutoff was chosen so that 
on average the 10\% slowest particles were used in the calculation. 
The choice of the cutoff
has little effect on the lifetime, as long as a sub-ensemble
with a relaxation time longer than the average relaxation time of the
full ensemble is identified.  The lifetime calculated by 
identifying the slow particles is always 
equal to the $\alpha$ relaxation time to within the uncertainty of the data.  

%Another measure of the mobility 
%which has been used in previous simulations \cite{KobDH,Donati} is
%used previously \cite{KobDH,Donati} is
Refs. \cite{KobDH,Donati} used the following measure of the mobility
\begin{equation}\label{d}
\delta_i(\Delta t) = \left | \mathbf{r}_i(t_2) - \mathbf{r}_i(t_1) \right |^2,
\end{equation} 
where $\Delta t = t_2 - t_1$.  We defined a slow sub-ensemble as the 
10\% with the smallest $\delta_i(3\tau_{\alpha})$, and calculated $F_{slow}^{\alpha}$ for this 
sub-ensemble.  Again, the average relaxation time of the sub-ensemble was longer than the average relaxation
time of the full ensemble.
The lifetime of the sub-ensemble defined using the second definition of the mobility, $\tau_{\delta}$,  
is equal to the $\alpha$ relaxation time to within the 
uncertainty of the data except for the A particles at the highest temperatures examined
in this work (see Fig.~\ref{lifetime}).

\begin{figure}[t]
	\includegraphics[scale=0.29]{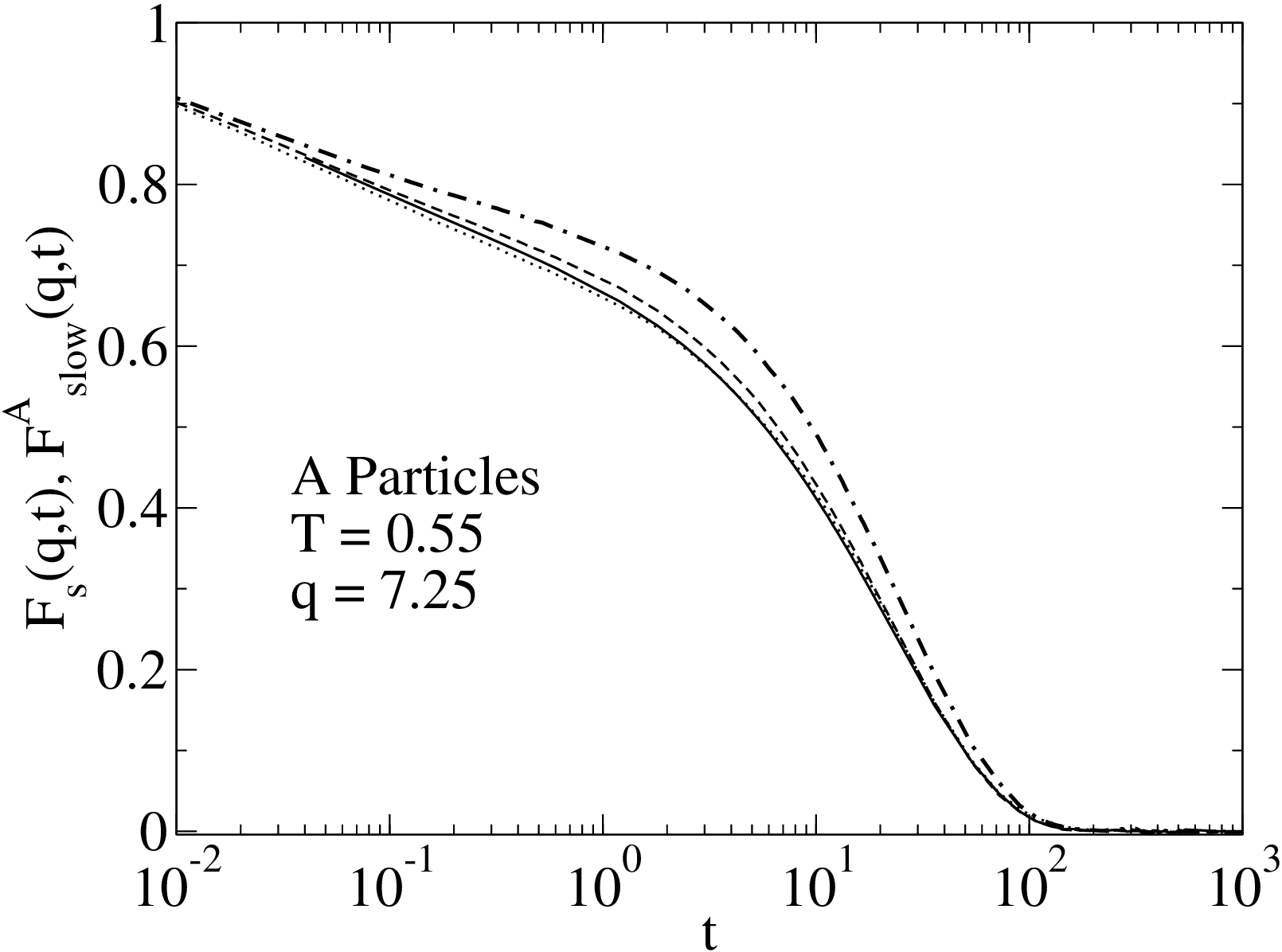}
\caption{$F_{slow}$ for different different sub sets of particles, 
$\mathcal{S} \cap \mathcal{D}$ (dashed dotted), $\mathcal{S} - \mathcal{D}$ (dotted),
$\mathcal{D} - \mathcal{S}$ (dashed), compared to $F_s^A$ (solid).  See text
for definition of sets $\mathcal{S}$ and $\mathcal{D}$.\label{compare}}
\end{figure}

To try to understand why both definitions give similar results, it is illustrative to examine the
relaxation of different subsets of particles chosen by the two definitions of mobility.  
Let $\mathcal{S}$ be the set of particles selected using $\sigma_i$ as the definition of
mobility, and $\mathcal{D}$ be the 
set of particles selected using $\delta_i$ as the definition of mobility.  
Figure \ref{compare} compares $F_{slow}^A$ for 
$\mathcal{S} \cap \mathcal{D}$, $\mathcal{S} - \mathcal{D}$, and $\mathcal{D} - \mathcal{S}$ 
to $F_s^A$ for $T = 0.55$.  The relaxation of the 
particles which are in set $\mathcal{S}$ but not $\mathcal{D}$,
or are in set $\mathcal{D}$ but not $\mathcal{S}$, is similar to the relaxation of the full
ensemble, but the particles which are in both sets have a longer relaxation time.  
Thus, the two definitions of 
mobility give similar results since they both are able to select 
%the essential sub-ensemble of slow particles, \textit{i.e.} the 
the particles whose average relaxation time is longer
than the average relaxation time of the full ensemble.  

%We also investigated a definition of slow particles used in Ref. \cite{Qianrate}.
%It resulted in $F_{slow}^{\alpha} \approx F_s^{\alpha}$. Thus, this definition 
%does not select a sub-ensemble of slow particles.

In conclusion, we used a four-time correlation function to select a slow 
sub-ensemble and analyze the dynamics of the slow sub-ensemble. 
The lifetime of the slow sub-ensemble selected by the four-time correlation
function is not longer than the $\alpha$ relaxation time. On approaching $T_c$
the lifetime increases faster with decreasing temperature than the $\alpha$ relaxation time \cite{comment}.
Closer to $T_c$ (beginning approximately at the temperature at which deviations from mode-coupling-like
power laws appear) the lifetime follows the temperature dependence of the $\alpha$ relaxation time.
We also identified two other slow sub-ensembles whose average relaxation time
is longer than the average relaxation time of the full ensemble using 
two different definitions of
mobility.  The essential sub-ensemble, the sub-ensemble chosen such that
$F_{slow}^{\alpha} \approx M^{\alpha}$, consists of the particles which stay closest to 
their position at $t_1$ over the time interval $\Delta t = t_2 - t_1$, and 
are still close to their position at $t_1$.  This suggests that
the slow sub-ensemble are the particles which are confined to their 
cage over the time interval $\Delta t$.
The lifetime of the slow sub-ensemble depended on the definition of mobility.  
If $\sigma_i$ was used to define mobility, the lifetime was equal to the $\alpha$ relaxation time 
at all temperatures.  If $\delta_i$ was used as the definition of mobility,
the lifetime was equal to the $\alpha$ relaxation time except for the $A$ particles at the 
highest temperature studied, in which case the lifetime was less than the $\alpha$ relaxation
time.  

Our findings qualitatively agree with NMR results of B\"ohmer \textit{et al.} \cite{Bohmer}.
Note, however, that there is a significant difference in the temperature of simulations
and experiments: the simulations have been performed slightly above $T_c$ whereas the experiments were done 
well below $T_c$. Thus, direct comparison of the two sets of results is impossible. The same
comment applies, however, to almost all simulational studies of glassy dynamics.

This work was supported by NSF Grant CHE 0111152.


\begin{thebibliography}{99}
\bibitem{comment}The $\alpha$ relaxation time and the lifetime of the slow sub-ensemble 
were both defined by the condition that an appropriately defined function decayed to an
arbitrary value of $e^{-1}$. 
If, instead of $e^{-1}$, 0.25, 0.35 or 0.5 were used,
the same trends were observed.
\end{thebibliography}
\end{document}